\documentclass[12pt]{article}
\usepackage{amssymb,amsmath,epsfig}

\begin{document}

\title{\bf Singularities of Noncompact Charged Objects}
\author{M. Sharif \thanks {msharif.math@pu.edu.pk} and G. Abbas
\thanks {abbasg91@yahoo.com}\\
Department of Mathematics, University of the Punjab,\\
Quaid-e-Azam Campus, Lahore-54590, Pakistan.}

\date{}

\maketitle
\begin{abstract}
We formulate a model of noncompact spherical charged objects in the
framework of noncommutative field theory. The Einstein-Maxwell field
equations are solved with charged anisotropic fluid. We choose the
forms of mass and charge densities which belong to two parameter
family of density distribution functions instead of densities as
Gaussian width length. It is found that the corresponding densities
and the Ricci scalar are singular at origin whereas the metric is
nonsingular indicating a spacelike singularity. The numerical
solution of the horizon equation implies that there are either two
or one or no horizon depending on the mass. We also evaluate the
Hawking temperature which implies that a black hole with two
horizons is evaporated to an extremal black hole with one horizon.
\end{abstract}

\section{Introduction}

Recently, there has been a growing interest to study the nature of
spacetime singularity and black holes (BHs) in the context of
noncommutative (NC) field theory. This is perhaps due to the fact
that some problems of the BH physics remain unanswered even after
passing many years \cite{1}. For example, the satisfactory
description of the final stage of BH evaporation is still an
unknown. According to BH correspondence principle \cite {2}, one
cannot neglect the string effects during the last stage of BH
evaporation. This has inspired the researchers to adopt the string
field theoretic approach in the various aspects of theoretical
physics. Noncommutative field theory is one of the outcomes of the
string theory in which spacetime coordinates become noncommuting
operators on a $D$-brane \cite{3},\cite{4}. This motivates the
researchers to reconsider the older ideas of Snyder \cite{5}.

Zade et al. \cite{5a} discussed the gravitational collapse of
radiating Dyon field and clarified the status of CCH. Nashed
\cite{5b} derived the regular charged solutions in Mull$\ddot{o}$r's
tetrad theory of gravitation. Ying et al. \cite{5c} formulated the
double NC form of a spacetime and introduced the complex symmetric
theory of gravitation. Guang \cite{5d} investigated the
singularities of static spherically symmetric charged scalar field
solutions

The spacetime noncommutativity can be expressed by the following
relation \cite{6}
\begin{equation}\label{6}
[x^\mu,x^\nu]=\iota{\sigma}^{\mu\nu},
\end{equation}
where ${\sigma}^{\mu\nu}$ is an anti-symmetric matrix that
determines the spacetime cell discretization as $\hbar$ (Planck's
constant) discretizes the phase space. There are several approaches
to formulate the NC field theory out of which one is based on the
$\star$-product and another on the coordinate coherent state
formalism. Using the second approach, Smailagic and Spallucci
\cite{7} explored that the problems of Lorentz invariance and
unitary arising in the $\star$-product can be solved by assuming
${\sigma}^{\mu\nu}={\sigma} diag(\epsilon_{ij},\epsilon_{ij}....)$,
where ${\sigma}$ is constant having dimensions of length squared.
Also, the coordinate coherent state modifies the Feynman
propagators. It is believed that NC would remove the singularities
(divergences) appearing in general relativity (GR). In GR, the
metric field is a geometrical structure and curvature measures its
strength. Since noncommutativity is the fundamental property of the
metric, so it affects the geometry through the field equations and
hence the energy-momentum tensor. In the NC GR, we take the geometry
part of the field equations unchanged, while a small change is
introduced in the matter part.

It was pointed out by Doplicer et al. \cite{8} that when matter
density is extremely large, a BH is formed. According to Heisenberg
uncertainty principle, the measurement of a spacetime separation
causes an uncertainty in momentum, i.e., momentum preserves inverse
proportionality to the extent of the separation. For small enough
separation, momentum becomes large and system leads to the BH
formation. The behavior of BH in the NC field theory has been
studied by many people. Banerjee et al. \cite{9} examined the
behavior of the NC Schwarzschild BH while Modesto and Nicolini
\cite{10} discussed the charged rotating NC BH. Bastos et al.
\cite{11},\cite{12} explored the non-canonical phase space,
singularity problem and BH in the context of NC geometry. Also,
Bartolami and Zarro \cite{13} investigated the NC correction to
pressure, particle numbers and energy density for fermion gas and
radiations. The NC correction to these quantities lead to the fact
that NC affects the matter dispersion relation and equation of
state. Inspired by the NC correction to BH physics, Oh and Park
\cite{14} explored the gravitational collapse of shell with smeared
gravitational source in the NC Schwarzschild geometry. We have
extended this work for NC Reissner-Nordstr$\ddot{o}$m background
\cite{15}. Sun et al. \cite{16} studied gravitational collapse of
spherically symmetric star in NC GR using spacetime quantization
approach.

In all these papers, the matter density is taken as Gaussian
distribution function. However, Castro \cite{17} studied the
singularity associated with noncompact matter source by using matter
density in a particular form which belongs to the most general two
parametric form of the density distribution. Here we extend this
work to study the nature of singularity associated with noncompact
charged matter sources extending from $r=0$ to $r=\infty$. The plan
of the paper is as follows: In the next section, we formulate the
noncompact charged object model. Section \textbf{3} is devoted to
discuss some properties of this model. In the last section, we
conclude our results.

\section{Noncompact Charged Objects Model}

To formulate noncompact charged objects model, we solve the
Einstein-Maxwell field equations with anisotropic fluid and
electromagnetic field. In the paper \cite{17}, instead of choosing
density in the form of Gaussian width length \cite{6}
\begin{equation}\label{2}
\rho=\frac{M_0e^{-\frac{r^2}{4\sigma^2}}}{(4\pi
\sigma^2)\frac{3}{2}},
\end{equation}
it was taken as
\begin{equation}\label{3}
\rho(r,\sigma)=\frac{M_0}{4\pi r^2}\frac{3\sigma^3}{2}
\frac{1}{r^4[1+(\frac{\sigma}{r})^3]^{\frac{3}{2}}},
\end{equation}
where $\sigma$ is NC parameter and the constant $M_{0}$ corresponds
to the total gravitational mass of the system.

The form of matter density $\rho$, given in Eq.(\ref{3}) belongs to
the most general two parameters family of the density distribution
function
\begin{equation}\label{4}
\rho(r,\sigma,k)=\frac{M_0}{4\pi r^2}\frac{k\sigma^k}{2}
\frac{1}{r^{1+k}[1+(\frac{\sigma}{r})^k]^{\frac{3}{2}}},\quad k>2.
\end{equation}
We follow \cite{17} and take the same expression for matter density
as in (3) and use the charge density in the form
\begin{equation}\label{5}
\rho(r,\sigma)=\frac{q_0}{4\pi r^2}\frac{3\sigma^3}{2}
\frac{1}{r^4[1+(\frac{\sigma}{r})^3]^{\frac{3}{2}}},
\end{equation}
instead of
\begin{equation}\label{6}
\rho=\frac{q_0e^{-\frac{r^2}{4\sigma^2}}}{(4\pi
\sigma^2)\frac{3}{2}},
\end{equation}
as given in \cite{18}. Here constant $q_{0}$ corresponds to the
total charge of the system.

Taking matter density $\rho$ and electric density $\rho_{el}$ as
defined in Eqs.(\ref{2}) and (\ref{6}), respectively, many authors
\cite{6},\cite{9},\cite{10},\cite{18} have derived the neutral as
well as charged and rotating NC BH solutions. All these exhibit a
de-Sitter core at the origin of the underlying geometry due to the
quantum fluctuations. Applying the same strategy as in above studies
but using a modified definition of the matter density (as defined in
Eq.(\ref{3})), Castro \cite{17} derived a NC solution which has no
de-Sitter core and exhibits the timelike naked singularity at
origin. The present letter is the extension of \cite{17} to the
electromagnetic field case.

We take a spherically symmetric spacetime in the following form
\begin{equation}\label{7}
ds^{2}=f(r)dt^{2}-\frac{1}{f(r)}dr^{2}-r^2(d{\theta}^{2}+\sin^2\theta
d{\phi}^2),
\end{equation}
where $f(r)$ is unknown function to be determined from the field
equations.

The matter under consideration is charged anisotropic fluid. The
components of anisotropic fluid energy-momentum tensor are
\begin{equation}\label{8}
T_{\mu}^{\mu}= (\rho(r), -p_r, -p_\theta, -p_{\phi}).
\end{equation}
The energy-momentum tensor of the electromagnetic field is
\begin{equation}\label{9}
{T_{\mu}^{\nu}}^{(em)}=\frac{1}{4{\pi}}
(-F^{{\nu}{\lambda}}F_{{\mu}{\lambda}}+\frac{1}{4}\delta^{\nu}_{\mu}
F_{{\pi}{\lambda}}F^{{\pi}{\lambda}}).
\end{equation}
The Maxwell equations are given by
\begin{eqnarray}\label{10}
F_{\mu\nu}=\phi_{\nu,\mu}-\phi_{\mu,\nu},\quad
{F^{\mu\nu}}_{;\nu}=4\pi J^{\mu},
\end{eqnarray}
where $F_{\mu\nu}$ is the Maxwell field tensor, $\phi_{\mu}$ is the
four potential and $J_{\mu}$ is the four current. For the static
spherically symmetric charge distribution, the four potential and
the four current are taken in the following form
\begin{equation}\label{12}
\phi_{\mu}=\phi{\delta^{0}_{\mu}},\quad J^{\mu}=\rho_{el}
{\delta^{\mu}_{0}},
\end{equation}
where $\phi$ is electric potential and $\rho_{el}$ is the electric
charge density. Solving the Maxwell field equations, we get
\begin{equation}\label{13}
F_{01}=-\frac{\partial \phi}{\partial r}=\frac{q(r,\sigma)}{r^2},
\end{equation}
where
\begin{equation}\label{14}
q(r,\sigma)=4{\pi}{\int^{r}_{0}}r^2 \rho _{el}dr= \frac
{q_0}{\sqrt{{1+(\frac{\sigma}{r})^3}}}.
\end{equation}

In order to find the relation between different components of the
energy-momentum tensor, we use the conservation of energy-momentum
tensor, i.e, $\tilde{T}^\mu _{\nu;\mu}=0~(\tilde{T}^\mu _\nu=T^\mu
_\nu +{T^{\mu}_{\nu}}^{(em)})$. For the Schwarzschild like nature of
the solution, i.e, $g_{00}=({g_{rr}})^{-1}$, we restrict that ${T}^0
_0={T}^r_r$, which implies that $p_r=-\rho(r)$ (negative radial
pressure pointing towards the center $r=0$) and $p_\theta=p_{\phi}$.
With these conditions, the conservation equation provides the
following relation
\begin{equation}\label{16}
p_\theta=-\frac{r}{2}
\partial_r(\rho+\frac{q^2}{8\pi r^4})-\rho-\frac{q^2}{4\pi r^4}.
\end{equation}
The Einstein field equations with Eqs.(\ref{3}),
(\ref{7})-(\ref{9}), (\ref{13}) and (\ref{16}) lead to the following
solution
\begin{eqnarray}\label{17}
ds^2 &=& (1-\frac{2m(r,\sigma)}{r}-\frac{Q^2(r,\sigma)}{r})dt^2-
(1-\frac{2m(r,\sigma)}{r}-\frac{Q^2(r,\sigma)}{r})^{-1}dr^2 \nonumber\\
&-& r^2(d{\theta}^{2}+\sin^2\theta d{\phi}^2),
\end{eqnarray}
where
\begin{equation}\label{18}
m(r,\sigma)= 4\pi{\int_0}^r \rho r^2 dr=\frac
{m_0}{\sqrt{{1+(\frac{\sigma}{r})^3}}}
\end{equation}
and
\begin{equation}\label{19}
{Q^2(r,\sigma)}=\frac{{q_0}^2}{18\sigma}{\left(\sqrt{3}\pi
-6\sqrt{3}\arctan(\frac{\sigma-2r}{\sqrt{3}\sigma})+log
\left(\frac{(\sigma^2-\sigma r+r^2)^3}{(\sigma+r)^6}\right)\right)}.
\end{equation}
The solution given in Eq.(\ref{17}) represents a noncompact charged
object.

\section{Properties of the Model}

This section is devoted to explore the properties of charged object
model. In the limit $\sigma\rightarrow0$, the line element
(\ref{17}) reduces to the classical Reissner-Nordstr$\ddot{o}$m
solution. We observe that the metric component
\begin{equation}\label{20}
g_{00}=1-\frac{2m(r,\sigma)}{r}-\frac{Q^2(r,\sigma)}{r}
\end{equation}
with Eqs.(\ref{18}) and (\ref{19}) satisfies the condition
${g_{tt}(r=0)}={g_{tt}}(r=\infty)=1$. This implies that the solution
(\ref{17}) is asymptotically flat and not singular at $r=0$. The
corresponding Ricci scalar is
\begin{equation}\label{21}
R= -\frac{3}{2}(\frac{\sigma}{r^3})^3
\left({5\sigma^3m+2r^2{q_0}^2\sqrt{{1+(\frac{\sigma}{r})^3}}-4mr^3}\right)
\end{equation}
which is singular at $r=0$. Also, the Ricci scalar can be found
directly from the field equations as $R=-T^\mu_\mu$, where we have
used ${T^\mu_\mu}^{(em)}=0$. Further, using the values of
$T^\mu_\mu$ along with $p_r$ and $p_\theta$, we get
\begin{equation}\label{22}
R=-\left(4\rho+\frac{r}
\partial_r(\rho+\frac{{q}^2}{8\pi r^4})+\frac{{q}^2}{4\pi
r^4}\right).
\end{equation}
After a straightforward but laborious calculations, we get the same
relation for $R$ as in (\ref{21}). This verifies the fact that the
solution (\ref{17}) is the valid solution of the Einstein field
equations with the charged anisotropic source. Despite the fact that
the metric is nonsingular at $r=0$, the Ricci scalar is singular at
$r=0$. This is due to the direct dependence of the Ricci scalar on
the matter density, which is singular at $r=0$. When $g_{00}(r)=0$
at the horizon radius $r=r_H$, then the metric (\ref{17}) becomes
singular, however, this is a coordinate singularity.

The horizon equation, $g_{00}(r)=0$, for the solution (\ref{17})
indicates that it is not possible to find exact solution of the
horizon equation for $r$. Thus with the arbitrary choice of
parameters, we evaluate the values of $r$ for $g_{00}(r)=0$ shown in
the left graph of Figure \textbf{1}. This shows that for $m_0=
0.44$, there exists one horizon (extremal black hole green curve),
when mass is smaller than this value, there exists no horizon (blue
curve) and two horizons exist (inner and outer horizons pink curve)
for mass larger than $m_0= 0.44$. Thus the case $m_0<0.44$
corresponds to the existence of naked singularity, as there do not
exist horizons to hide the singularity at $r=0$. Thus the minimum
mass for the existence of BH in this framework is $m_0= 0.44$. Since
matter density as well as the Ricci scalar diverge at $r=0$, while
the metric components are finite there so singularity is spacelike.
Thus, we find that the solution represents a spacelike naked
singularity, extremal BH and a BH with inner and outer horizons
depending on the mass.

The Hawking temperature for the BH is
\begin{equation}\label{24}
T_H=\left(\frac{1}{4\pi}\frac{dg_{00}}{dr}\right)_{r=r_H}.
\end{equation}
The behavior of $T_H$ (after eliminating $m=m(r,q_0,\sigma)$ from
$\frac{dg_{00}}{dr}$ and $g_{00}(r)=0$) is shown in the right graph
of Figure \textbf{ 1}. As $r_H$ decreases, temperature increases and
attains a maximum value at $r_H\simeq2$, then suddenly drops to zero
at $r_H\simeq1.1$ corresponding to the radius of the extremal BH.
Since further decrease in $r_H$ results to negative temperature
which is not acceptable. Thus there does not exist naked singularity
at $r_H=0$ in this case. Consequently, singularity at $r=0$ is
covered by extremal horizons, hence the possibility of mass less
than critical mass is excluded from the discussion.
\begin{figure}
\center\epsfig{file=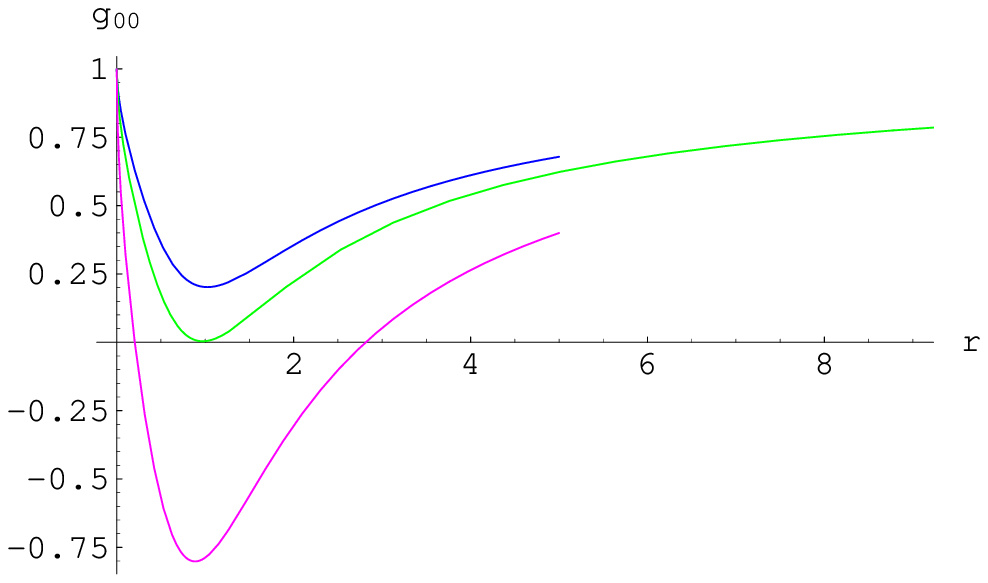, width=0.35\linewidth} \epsfig{file=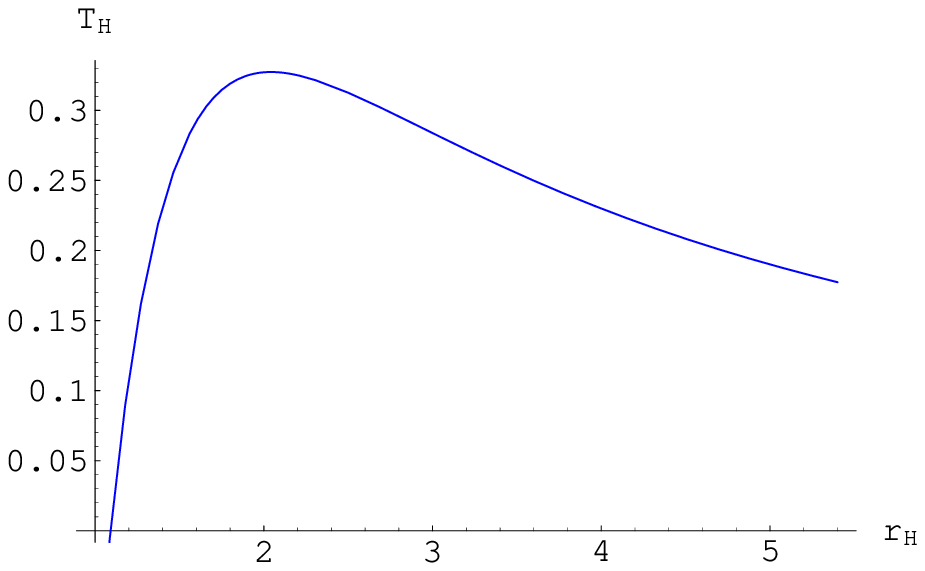,
width=0.35\linewidth}\caption{The left graph is $g_{00}$ versus
$r,~q_0=1,~\sigma=1$, the intercepts on the horizontal axis give the
radius of event horizons. For $m_0=1$ (pink curve) two horizons;
$m_0=0.44$ (green curve) extremal BH and $m_0=0.3$ (blue curve) no
horizon. The right graph shows that $T_H$ versus $r_H$. In this
graph $T_H=0$ for $r_H\backsimeq1.1$, i.e., for the extremal BH,
while $T_H\backsimeq 0.32$ corresponds to mass $m\backsimeq0.6$}
\end{figure}

The radial null geodesics from the singularity at $r=0$ reach to an
observer in the finite (naked singularity case) and infinite (one
and two horizons cases) time. The equation of the radial null
geodesics is
\begin{equation}\label{24}
ds^2=(1-\frac{2m(r,\sigma)}{r}-\frac{Q^2(r,\sigma)}{r})dt^2
-(1-\frac{2m(r,\sigma)}{r}-\frac{Q^2(r,\sigma)}{r})^{-1}dr^2=0,
\end{equation}
which can be written as
\begin{equation}\label{25}
t=\int^r_0
\frac{dr}{1-\frac{2m(r,\sigma)}{r}-\frac{Q^2(r,\sigma)}{r}}.
\end{equation}
For one and two horizon cases, i.e., $m_0=0.44$ and $m_0>0.44$,
respectively, the graph shows that the integral diverges. In both
cases, we get $t=\infty$ for $r\neq\infty$. This implies that the
light signal coming from the singularity along the null radial
geodesics take infinite time to reach the distant observer, these
never reach to the observer at $r\neq\infty$.

\section{Conclusion}

In this letter, we have formulated a new static, spherically
symmetric charged solution of the Einstein-Maxwell field equations
in the framework of NC geometry. This describes the final stage of
the collapsing charged noncompact object. This work extends the work
of Castro \cite{17} to the charge case and its results can be
recovered by taking the charged parameter $Q=0$. The present
solution reduces to the classical RN solution, by taking the NC
parameter $\sigma\rightarrow0$. It is found that the solution is
asymptotically flat and not singular at the origin, while the Ricci
scalar and matter density are singular, hence spacelike singularity
appears in this case. The horizons analysis implies that for a
suitable set of initial data and $m_0<0.44,~m_0=0.44$ and
$m_0>0.44$, there exist no horizon, single horizon and two horizons
respectively. However, the thermodynamical results imply that the
final stage of the object is extremal BH containing a singularity at
$r=0$.

In the paper \cite{17}, the horizons equation is cubic polynomial
which admits three possible exact solutions that correspond to two
horizons, one horizon and no horizon cases. It was concluded on the
basis of Hawking temperature (with the assumption that given matter
configuration has mass less than the critical mass) that there
exists timelike naked singularity. In our case, we conclude that
spacelike singularities associated with the noncompact charged
object are covered by extremal BH horizon. There is curvature
singularity at $r=0$, while there is de-Sitter geometry around the
origin \cite{6} instead of curvature singularity.

\vspace{1.50cm}

{\bf Acknowledgment}

\vspace{0.25cm}

We would like to thank the Higher Education Commission, Islamabad,
Pakistan for its financial support through the {\it Indigenous Ph.D.
5000 Fellowship Program Batch-IV}.

\end{document}